\documentclass[12pt]{article}

\begin{document}

\begin{center}

\section*{Finite-Size Scaling  and Long-Range  Interactions}

\vskip 5mm N.S. Tonchev

 \vskip 5mm

 {{\it Inst. of Solid State Physics, Bulg. Acad. of Sci.,
Tzarigradsko shauss$\acute{e}$e 72, 1784 Sofia, Bulgaria.}
\\

}
\end{center}
\vskip 5mm

\begin{abstract}
The present review is devoted to the problems of finite-size
scaling due to the presence of long-range interaction decaying at
large distance as $1/r^{d+\sigma}$, where $d$ is the spatial
dimension and the long-range parameter $\sigma>0$. Classical and
quantum systems are considered.
\end{abstract}
\vskip 8mm

\subsection*{1. Finite-size systems and critical phenomena}

  A common wisdom is that the singularities
in the thermodynamic functions at a critical point may occur only
in the  thermodynamic limit. If the system is fully finite (as any
real system is), no such singularities exist and, strictly
speaking, no phase transitions occur. These facts naturally give
rise to the following questions:

a) Why the bulk theory turns out to be adequate to the
experimental evidence for finite  objects?

b) How the singularities appear  when no phase transitions occur
in any fully finite system?

The answer to the first question lies in the fact that a
macroscopic body is close, under some circumstances, to the
idealized thermodynamic limit.

The answer to the second question becomes evident if one recalls
that the limit function of a sequence of analytic functions needs
not to be analytic. Therefore, what happens is that in the
thermodynamic limit some thermodynamic functions of the system
develop singularities which are attributes of phase transitions.
So, in a finite system one expects an appreciable rounding of the
critical point singularities.

 The bulk correlation length $\xi_{\infty}(T)$
measures the distance over which particles in a system are
significantly correlated. Then, as the temperature $T$ approaches
the critical temperature $T_c$, $\xi_{\infty}(T)$ diverges. When
$\xi_{\infty}(T)$ attains a magnitude of the order of the
characteristic size $L$ of the finite system, then the boundary
particles at the opposite sides of the system become well
correlated, and ordering cannot continue to build up further in
the restricted dimensions. It is certainly reasonable to expect
that the rounding of the phase transition is controlled by the
criterion: $\xi_{\infty} (T)\sim L.$

The description of this rounding and crossover was first
formulated by M. Fisher (1972) and is called Finite-Size Scaling
(FSS)( see e.g.\cite{brankov00}).

An important application of FSS is to analyze numerical data
obtained from simulations on relatively small finite systems, and,
thereby to obtain knowledge of the bulk systems.

Finite-size effects are much  stronger in systems of particles
with long-range (LR) interactions because every particle directly
feels the influence of the boundaries.

 To study finite systems, one must at the start address two basic
issues, namely the overall geometry of the system and the specific
nature of the boundary conditions.

\textsc{Geometry}: We will use the general notation
 $L^{d-d'}\times\infty^{d'}$ and distinguish
three important cases ($d$,$d'$ are spatial dimensions): $ d'=0$
-- a fully finite system,
 $d'=1$ --
  a system with the geometry of a cylinder,
and $d'=(d-1)$ --
 a system with slab (or film) geometry.

\textsc{Boundary conditions}: Here and bellow we will consider
periodic boundary conditions (p.b.c.). Close to the bulk critical
point
 for a $d$-dimensional system with a finite
size $L$ one has for the free energy density
\begin{equation}
\label{perbc}
 f_{L}^{(p)}(T) = f(T)+ O\left({\mathrm e}^{-L/\xi (T)}\right).
\end{equation}
where $f(T)\equiv f_{\infty}(T)$ is the bulk free energy density.
In this case there are no surfaces, edges and corners.

The experimental observability of FSS effects is hampered by the
fact that nature does not provide us systems with p.b.c. The rapid
exponential approach to the thermodynamic limit is one of the
reasons why p.b.c. are preferred in Monte Carlo simulations. One
observes that the l.h.s. in the above expansions  is a regular
function of $T$  whereas the r.h.s. is not: the bulk density
$f(T)$, for example, is singular at the critical point $T_c$.
Therefore, Eq.(\ref{perbc}) can hold only {\it away} from $T_c$.
Obviously, for $T\simeq T_c$ one needs an alternative,
 (FSS) formulation.

\subsection*{2. Finite-size scaling hypothesis} Scaling hypotheses are
made in the form of  statements about properties of
 thermodynamic quantities in terms of homogeneous functions.

 To simplify the further
discussion we consider that: the studied system is below its upper
critical dimension $d_{u}$, the infinite  system has a second
order phase transition
 at a critical temperature $T_{c}$, and the  system has length $L$
in all finite directions. The violation of these restrictions will
cause complications and
 modifications of the standard FSS hypothesis \cite{brankov00}.

 For a
 physical quantity $\textit{P}$ with singular behavior in the
thermodynamic limit the content of the standard FSS hypothesis is
to assume the existence of a scaling function $\textit{P}_{L}(T)$
such that:
\begin{equation}\label{FSS1}
\textit{P}_{L}(T)\simeq\textit{P}_{\infty}(T)\textit{F}_{L}(L/\xi(T)).
\end{equation}
As far as $L$ is finite, $\textit{P}_{L}(T)$ must be a regular
function of $T$. The (universal) function $\textit{F}_{L}(x)$ must
compensate the singularity of $\textit{P}_{\infty}(x)$.
Eq.(\ref{FSS1}) may be written in equivalent form:


\begin{equation}\label{FSS2}
 P_{L}(t)\simeq L^{p_x/\nu}{\cal F}_{L}(tL^{1/\nu}),
\end{equation}
where  $p_x$ - the critical exponent of the observable $ P_{L}$,
$\nu$  - the critical exponent of the correlation length
 and $t=(T-T_{c})/T_{c}$.
The hypothesis is expected to be valid when $L$ is large and $T$
approaches $T_{c}$.

 What is the common status of the above hypothesis? One can usually meet
 the following statements,
e.g. J. Cardy, in \cite{cardy96}:
 "FSS  is by now well-established theoretically, at least for
systems with short-range interactions ...", or  quite recently
 Chen
and Dohm in \cite{CD02} :
 "The finite-size scaling hypothesis asserts ... in \textit{the absence} of
LR  interaction ..."

We would like to point out the restrictions  related with the
nature of the interaction - one insists that  it must be
short-range (SR). In this situation the problem is to find out:
 does the FSS take place if the interaction is of the
LR type? As we will explain bellow there is a positive answer to
this question.

We will consider interactions which decay algebraically at large
distances with the standard notation

\begin{equation}\label{01}
u(r)\sim-1/r^{d+\sigma},\qquad\sigma>0.
\end{equation}

For $\sigma\leq0$ the interactions are \textit{nonintegrable}, and
so, under standard definitions, the thermodynamic limit does not
exist. The case of \textit{nonintegrable} interaction is beyond
the scope of the present study.

The LR interaction (\ref{01}) enters the expressions of the theory
only through its Fourier transform \cite{brankov00}. The
corresponding small - ${\mathbf q}$ expansion of the Fourier
transform has the general form
\begin{equation}\label{1}
v({\mathbf q}) = v_{0} + v_{2}{\mathbf q}^{2} +v_{\sigma}
{|\mathbf q|}^{\sigma}+w({\mathbf q}) \qquad  0< \sigma \neq 2,
\end{equation}
with $w({\mathbf q})/{|\mathbf q|}^\sigma\to 0$, for ${\mathbf
q}\to 0$,  i.e. in the long-wavelength approximation. Further on
we will formally relate $\sigma=2$, to the SR  interaction since
then~(\ref{1}) is the Fourier transform of an interaction decaying
exponentially with distance.

Depending on whether in~(\ref{1}) $\sigma $ is less or bigger than
2 we will speak about {\it leading} or {\it subleading} LR term
respectively.

\textsc{Nota bene}: The \textit{subleading} LR term does not
affect the bulk critical behavior, but influences the finite-size
one (for details see \cite{dantchev01,chamati2001,D01}).

 The defined above LR interaction has
to be a mimicry of the real one. Nevertheless, from the pure
theoretical point of view the considered LR interaction seems to
be of specific interest. The reason is that renormalization group
 (RG) predictions obtained on the basis of the
$\epsilon$-expansion, can be verified on ideal testing ground. The
upper and lower critical dimensions are $2\sigma$ and $\sigma$
respectively. Since $\sigma$ is a continuous parameter
 the value of $\epsilon=2\sigma-d$ or $\epsilon=d-\sigma$
would be small enough for integer values of the dimensionality.
This  places us in the rare situation when the outcome of the
computer simulations, obtained by means of the Monte Carlo method,
can be directly compared to the
 predictions of the $\epsilon$-expansion with $\epsilon$ arbitrarily small.

 But along this line of consideration there are some problems.

 Firstly, although the general scheme  of the $\epsilon$-expansion has been widely
 accepted, the part of it related to LR systems has not been well understood
until now (see \cite{CT03} and references therein).

Secondly, the study of LR systems leads to prohibitively large
computational requirements. Still recently a novel Monte Carlo
algorithm which has an efficiency that is independent of the
number of interactions per  spin has been announced
\cite{LB95,LB97}.

Thirdly, Monte Carlo analyzes of the RG predictions typically
apply FSS concepts. However, specific problems arise in the proper
generalization of the FSS concepts in the case under consideration
\cite{BD91,BT92}.

\subsection*{3. Problems with the correct definitions of the
correlation length}

By definition the correlation length is
\begin{equation}
\xi_{1}(T)=-\lim_{R\rightarrow\infty}[R/\ln
G_{\infty}(\mathbf{R};T)].
\end{equation}
Alternatively, one may consider the second moment of the bulk pair
correlation function and define the effective correlation radius:
\begin{equation}
\xi_{2}(T)=\left[\sum_{\mathbf{R}}R^{2}G_{\infty}(\mathbf{R};T)/G_{\infty}(\mathbf{R};T)
\right]^{1/2}.
\end{equation}

These two most commonly used definitions of the bulk correlation
length are unambiguous and equivalent (up to a constant) in the
case of exponential with the distance $R=|\mathbf{R}|$ decay of
the bulk pair correlation function $G_{\infty}(\mathbf{R};T)$. A
distinctive feature of the LR interaction is that the function
$G_{\infty}(\mathbf{R};T)$ decays as $R^{-d+\sigma}$ when
$R\rightarrow\infty$ for $T>T_{c}$,  and both definitions yield
$\xi_{1}(T)=\xi_{2}(T)=\infty $ if $\sigma<2$.

Anyhow complications with correlations that decay as power laws in
a whole domain of thermodynamic parameters, rather than only at
special points, arise even in the bulk case, provoking the idea of
"generic scale invariance"
 in the theory of phase transitions (see, e.g.
\cite{BKV04}). The FSS under consideration encounters the same
phenomena mixed with specific boundary effects. One can overcome
the difficulty with the absence of good definition of the
correlation length (when $\sigma<2$) following the ideas proposed
in \cite{BD91,BT92}, where instead of $\xi(T)$ a bulk
\textit{characteristic length} $\lambda(T)$ is used. This
characteristic length determines the length-scale of variation of
the correlation function and diverges at the critical point.

A finite system has three characteristic length scales: a linear
size $L$, a length $\lambda_{L}(T)$ which determines the scale of
variation of the correlations, and a microscopic length $a$ (e.g.,
$a$ is the lattice spacing). The approach proposed in \cite {BT92}
 involves the following steps:

 \textsc{First step}: It consists of three assumptions.

The first one defines the finite-size characteristic length
$\lambda_{L}(T)$ from the large-distance asymptotic behavior of
the finite-size pair correlation function $G_{L} ({\bf R},T)$.

The second assumption is the homogeneity of any thermodynamic
quantity of the finite-size system, such that its bulk limit is singular at
$T = T_{c}$, as a function of the two dimensionless ratios $\lambda_{L}(T)/a$
and $L/a$.

The third assumption is the existence of finite thermodynamic limits
for the characteristic length and the pair correlation
function.

\textsc{Second step}: It concerns the relationship between the
finite-size length\index{characteristic length!finite-size|(}
$\lambda_{L}(T)$ and its bulk limit\index{characteristic
length!bulk} $\lambda_{\infty}(T)$; the corresponding homogeneity
assumption  involves the dimensionless ratios
$\lambda_{\infty}(T)/a$ and $L/a$.

\subsection*{4. Mathematical problems}\label{app}

\textsc{Classical case}:  In the case of SR interaction the
following replacement is used as an indispensable part of the FSS
calculations  (see, e.g. \cite{brankov00}).

\begin{equation}\label{Scr}
\sum_{\mathbf{q}}\frac{1}{m^{2}+|\mathbf{q}|^{2}}=\int_{0}^{\infty}dt\exp(-m^{2}t)
\left[\sum_{q}\exp(-q^{2}t)\right]^{d},
\end{equation}
where $\mathbf{q}$ and $q$ are d-dimensional and one-dimensional
discrete vectors, respectively. This is the so called Schwinger
parametric representation. The aim of this replacement is
two-fold: to reduce the $d$-dimensional sum to  the corresponding
effective one-dimensional one, and  to give  the dimensionality
$d$ the status of a continuous variable.

In the case of leading  $q^{\sigma}$ term  in (\ref{1}), one
cannot  use the Schwinger representation in its familiar form.
Then the following generalization of (\ref{Scr}) has been
suggested \cite{BT88}:
\begin{equation}\label{BTr}
\sum_{\mathbf{q}}\frac{1}{m^{2}+|\mathbf{q}|^{\sigma}}=
m^{\frac{4-2\sigma}{\sigma}} \int_{0}^{\infty}dt
Q_{\sigma}(m^{4/\sigma}t)\left[\sum_{q}\exp(-q^{2}t)\right]^{d},
\end{equation}
where the function $Q_{\sigma}(t)$ for $0<\sigma<2$ is given by
\begin{equation}\label{F}
Q_{\sigma}(t)= \int_{0}^{\infty}dy\exp(-ty){\tilde
Q}_{\sigma}(y),\qquad {\tilde
Q}_{\sigma}(y)=\frac{1}{\pi}\frac{\sin(\sigma\pi/2)y^{\sigma/2}}
{1+2y^{\sigma/2}\cos(\sigma\pi/2)+y^{\sigma}}.
\end{equation}

>From (\ref{BTr}) and (\ref{F}) one
obtains
\begin{equation}\label{int1}
\sum_{\mathbf{q}}\frac{1}{m^{2}+|\mathbf{q}|^{\sigma}}= \int_{0}^{\infty}dt{\tilde
Q}_{\sigma}\left(\frac{t}{m^{\frac{4-2\sigma}{\sigma}}}\right)
\sum_{\mathbf{q}}\frac{1}{tm^{2}+|\mathbf{q}|^{2}}.
\end{equation}
The identity (\ref{int1}) demonstrates the possibility to  reduce
the LR case  to the SR case with an integration over an additional
parameter. So, all the mathematical  machinery developed for the
SR case, in principle, may be used. Let us note however that such
nonlocal expressions are unsuitable for explicit calculations. It
is more convenient to use the relation \cite{B89}:

\begin{equation}\label{MLFa}
Q_{\sigma}(x)=x^{\sigma/2 -
1}E_{\sigma/2,\sigma/2}(-x^{\sigma/2}),
\end{equation}
where $E_{\alpha,\beta}(z)$
 is an entire
function of the Mittag-Leffler type defined by the power series
\begin{equation}\label{defMLa}
E_{\alpha,\beta}(z)=\sum_{k=0}^{\infty}\frac{z^{k}}{\Gamma(\alpha
k+\beta)} \qquad \alpha>0.
\end{equation}
 and to study the finite-size properties of the system resulting from the analytical properties
 of $E_{\alpha,\beta}$.

\textsc{Pure quantum case}: At $T=0$, the quantum counterpart of
(\ref{BTr}) is

\begin{equation}\label{qScr1}
\sum_{\mathbf{q}}\frac{1}{\left[m^{2}+|\mathbf{q}|^{\sigma}\right]^{1/2}}=
\frac{2}{\pi}\int_{0}^{\infty}dp\sum_{\mathbf{q}}\frac{1}{m^{2}+p^{2}+|\mathbf{q}|^{\sigma}},
\end{equation}

 Equation (\ref{qScr1}) displays the well known
property of the pure quantum case. The auxiliary variable $p^{2}$
acts effectively as an extra dimension. Indeed the pure quantum
system corresponds to a $d+1$ dimensional classical system with
the geometry of a cylinder $L^{d}\times\infty^{1}$. Introducing
$m^{2}(p):=m^{2}+p^{2}$ with the help of Eq.(\ref{int1}) from
Eq.(\ref{qScr1}) one gets

\begin{equation}\label{qScr11}
\sum_{\mathbf{q}}\frac{1}{\left[m^{2}+|\mathbf{q}|^{\sigma}\right]^{1/2}}=
\frac{2}{\pi}\int_{0}^{\infty}dt\int_{0}^{\infty}dp{\tilde
Q}_{\sigma}\left(\frac{t}{m(p)^{\frac{4-2\sigma}{\sigma}}}\right)
\sum_{\mathbf{q}}\frac{1}{tm(p)^{2}+|\mathbf{q}|^{2}}.
\end{equation}
The integral representation (\ref{qScr11}) illustrates  the
mathematical difficulties  appearing in the pure quantum case. It
is shown that by two additional integrations the problem can be
effectively reduced to the classical SR case.

On the other hand, the following modification of (\ref{BTr}) has
been prosed \cite{C94}
\begin{equation}\label{BTrH}
\sum_{\mathbf{q}}\frac{1}{\left[m^{2}+|\mathbf{q}|^{\sigma}\right]^{1/2}}=
m^{\frac{4-\sigma}{\sigma}} \int_{0}^{\infty}dt
C_{\sigma}(m^{4/\sigma}t)\left[\sum_{q}\exp(-q^{2}t)\right]^{d},
\end{equation}
where
\begin{equation}\label{MLF}
C_{\sigma}(x)=x^{\sigma/4 -1}G_{\sigma/2,\sigma/4}(-x^{\sigma/2}),
\end{equation}
i.e. the function $C_{\sigma}(x)$ is related  to the entire
function defined by the power series
\begin{equation}\label{defML}
G_{\alpha,\beta}(z)=\frac{1}{\pi^{1/2}}\sum_{k=0}^{\infty}\frac{\Gamma(
k+1/2)z^{k}}{\Gamma(\alpha
k+\beta)k!} \qquad \alpha>0,\beta>0.
\end{equation}

Some analytical properties of $G_{\sigma/2,\sigma/4}(z)$ (see also
\cite{CT2000})  may be established with the help of the identity
\begin{equation}
G_{\sigma/2,\sigma/4}(z)=\frac{2}{\pi}\int_{0}^{\infty}E_{\sigma/2,\sigma/2}
\left(-(z+p^{2})\right)dp,
\end{equation}
obtained  from Eq.(\ref{qScr1}) and the relation of its l.h.s and
r.h.s. with the functions $G_{\sigma/2,\sigma/4}(z)$ and
$E_{\sigma/2,\sigma/2}(z)$, respectively.
 \subsection*{5. Theoretical models}

Here we will mention the models and the related works in which
leading LR interactions are considered in the context of FSS.

The mean spherical model  was considered in
\cite{BD91,BT92,BT88,B89}, \cite{FP86}-\cite{CD04}. The
Hamiltonian of the model is
\begin{equation}\label{model1}
{\cal H}=-\frac12\sum_{ij}{\cal J}_{ij}\ {\cal S}_i{\cal
S}_j-H\sum_i{\cal S}_i,
\end{equation}
where ${\cal S}_i$ is the spin variable at site $i$, ${\cal
J}_{ij}$ is the interaction matrix between spins at sites $i$ and
$j$, and $H$ is an ordering external magnetic field. The spins
 obey the
spherical constraint $\sum_i\langle{\cal S}_i^2\rangle=N,$ where
$\langle\cdots\rangle$ denotes the standard thermodynamic average
taken with the Hamiltonian ${\cal H}$ and $N$ is the the total
number of spins.

 The quantum version of model (\ref{model1})  was considered in \cite{CDT00}.
 The Hamiltonian of the model is
\begin{equation}\label{modelq}
{\cal H}=\frac{g}{2}\sum_{i}{\cal P}_{i}^{2}-\frac12\sum_{ij}{\cal J}_{ij}\ {\cal S}_i{\cal
S}_j +\frac{\mu}{2}\sum_i{\cal S}_i^{2}-H\sum_i{\cal S}_i,
\end{equation}
where ${\cal S}_i$ are spin operators at site i, the operators
${\cal P}_{i}$ are "conjugated" momenta and $[{\cal S}_i,{\cal
S}_i']=i\delta_{ii'},[{\cal P}_{i},{\cal P}_{i'}]=i\delta_{ii'}$
(with $\hbar=1$). Model (\ref{modelq}) in the large-n limit is
equivalent to the quantum ${\cal O}(n)$ nonlinear sigma model.

The $d$-dimensional ${\cal O}(n)$-symmetric model  was considered
in \cite{KT91}-\cite{LB02}. The model is defined by
\begin{equation}\label{modela}
{\cal H}\left\{\varphi\right\}=\frac12\int_V d^dx\left[
\left(\nabla^{\sigma/2}\varphi\right)^2+r_0\varphi^2
+\frac12u_0\varphi^4\right],
\end{equation}
where $\varphi$ is a shorthand notation for the space dependent
$n$-component field $\varphi(x)$, $r_0=r_{0c}+t_0$ ($t_0\propto
T-T_c$) and $u_0$ are model constants. The first term  denotes
$k^\sigma|\varphi(k)|^2$ in the momentum representation. A
strongly anisotropic version of (\ref{modela}) with exponents
depending on the direction was considered in \cite{CGGP}.

The quantum version of model (\ref{modela}) is
\begin{equation}\label{modelq}
{\cal H}\left\{\varphi\right\}=\frac12\int_{0}^{1/T}d\tau\int_V
d^dx\left[ \left(\nabla^{\sigma/2}\varphi\right)^2+r_0\varphi^2
+\frac12u_0\varphi^4\right],
\end{equation}
where $\varphi$ is a shorthand notation for the space-time
dependent $n$-component field $\varphi(x,\tau)$ \cite{CT2000}.

\subsection*{6. Verification of FSS }
\subsubsection*{Classical phase transitions}

Analytically the case  $\sigma<d<2\sigma$ has been investigated
first exactly in the spherical limit ($n=\infty$)
\cite{BD91,BT92,BT88,B89}, \cite{FP86}-\cite{BT} and by the RG
methods for {$n=1$} \cite{luijten99}, $n\geq 1$
\cite{chamati01,chamati02}. Finite-size effects are predicted in
both  regions of first order \cite{FP86,SP89,BD91,BT92} and second
order phase transitions
\cite{BT88,BT9,B89,BT92,luijten99,chamati01,chamati02}. As a
result in the latter case it has been shown that the
  FSS   relations can be written in  form
 equivalent to Eq.(\ref{FSS2}).
Some thermodynamic quantities, for instance, the shift of the
critical temperature due to the finite-size effects \cite{CT96},
the susceptibility \cite{BT,BT9}, the pair correlation function
\cite{BD91,SP89} and the Binder's cumulant at the bulk critical
temperature $T_c$ \cite{luijten99,chamati01,chamati02} and above
it \cite{chamati01,chamati02} as a function of $ \sqrt{\epsilon}$
up to ${\cal O}(\epsilon^{3/2})$ have been obtained.

Away from the critical region, the critical behavior of the
systems is dominated by its bulk critical behavior

 \textsc{Nota bene}: A distinctive feature of the LR case is that for
$tL^{1/\nu}\gg1$ the finite-size corrections are not exponentially
small as in the SR case; they vary instead as algebraic power of
the variable $tL^{1/\nu}$ \cite{SP89,BD91,chamati01}.

Some problems exist with the comparison of RG analytical results
and numerical data. It has been shown \cite{luijten99} (see also
\cite{luijten2001}) using Monte Carlo simulation (for $n=1$), that
the Binder's cumulant ratio is linear in  $\epsilon$. The
analytical evaluation of the Binder's cumulant for the ${\cal
O}(n)$ symmetric $\varphi^4$ model, however, showed that it is
linear in $\sqrt\epsilon$. A possible way to resolve this
controversy between the Monte Carlo method and the analytical
results is to carry on finite-size calculations to higher loop
order \cite{luijten99}. This could ameliorate the analytical
results, which would be comparable to those obtained by numerical
simulations. Here we would like to mention that higher loop
corrections that are dealt with through minimal subtraction scheme
(see \cite{BZ85a,luijten99,chamati01}) and $\epsilon$-expansion
have not been performed, even for the more simple case of SR
interaction and therefore much knowledge is still absent.

\subsubsection*{Quantum phase transitions}

Quantum phase transitions \cite{SS} occur at zero temperature as a
function of some \textit{nonthermal} control parameter such as
composition or pressure and are driven by quantum fluctuations. In
addition to being of an interest for various low-temperature
phenomena, quantum phase transitions are important because they
are believed to play a crucial role in quantum information
science, e.g. in the phenomenon known as entanglement - the
resource that enables quantum computation and communications (see
\cite{HNO} and refs. therein).

The main feature of quantum systems is the coupling between
statics and dynamics. As a result the dynamical critical exponent
$z$ plays an important role in scaling of the "imaginary time
dimension". For nonzero temperature this extra "dimension" extends
only over a finite interval $L_{\tau}$. Here $L_{\tau}\sim1/T$ is
"finite-size" in the temporal (imaginary time) direction and so
the temperature's influence in the quantum critical point my be
determined by FSS effects \cite{brankov00}.

Unlike classical models, where scaling can be done uniformly for
all ``spatial'' dimensions, quantum models are anisotropic in
general, and therefore the ``space'' and ``imaginary-time''
directions will not scale in the same fashion. According to the
general hypothesis of the FSS theory extended here to quantum
(anisotropic) systems, a physical quantity ${\cal A}
\left(r,L,T\right)$, which is  singular  in the thermodynamic
(bulk) limit at the quantum critical point ($r=0$), will scale
like
\begin{equation}\label{FSSA}
{\cal A}(r,T,L)=b^p {\cal
A}_s\left(rb^{1/\nu},Tb^z,bL^{-1}\right).
\end{equation}
Here, $r$ measures the distance from the quantum critical point.
In the scaling form~(\ref{FSSA}) $p$ corresponds to the
engineering dimension $d+z$ of the system in the case when the
scaling function refers to the singular part of the free energy.
For the other physical quantities of interest it is   the critical
exponent measuring the divergence of the bulk thermodynamic
function ${\cal A}$ at the critical point divided by $\nu$ (for
the correlation length $p=1$, for the susceptibility
$p=\gamma/\nu$, etc). Depending on the choice of the RG rescaling
factor $b$ we obtain different scaling functions ${\cal A}_s$,
which are related among each other by some appropriate change of
the scaling variables \cite{CT2000}.

 The fact that the
inverse temperature can be used as an additional size in the
imaginary-time direction creates  strong anisotropy in the system
and in general, one needs a shape--dependent formulation  of the
FSS \cite{CT2000,CGGP}. This  will lead to the establishment of
some changes in the scaling properties of the finite quantum
system. In this  case we can consider the {\it quantum to
classical} and the {\it finite-size to the bulk} system
\cite{CT2000}, and as well as other \cite{CGGP} crossover
phenomena.

\subsubsection*{Results on the Casimir effect}

The correlation function serves as a measure of how fluctuations
at one point are correlated with fluctuations at another point.
The confinement conditions, imposed on a system with correlation
function decaying as \textit{power} law in space induce a LR force
between the surfaces limiting the system. One can generally call
this phenomenon Casimir effect. There can be different mechanisms
 leading to this phenomenon that are related with the
 scale invariance or mentioned above "generic scale invariance".
 Prominent examples
are systems at critical points or systems with spontaneously
broken global continuous symmetry that leads to massless modes:
"spine waves" or Goldstone bosons (statistical-mechanical Casimir
effect \cite{FG78}), and  the discussed for the first time by
H.G.B. Casimir (1948) constrained zero-point vacuum fluctuations
of the electromagnetic field (quantum - mechanical Casimir effect
).

 In a system with geometry $L^{1}\times
\infty^{d-1}$ at or bellow the critical point a LR
  ``statistical-mechanical Casimir
force'' \cite{brankov00,FG78,krech94} appears. According to the
standard FSS hypothesis, near the critical point of the bulk
system one expects for this force  (the magnetic field H=0)
\begin{equation}\label{CFH}
F_{{\rm Casimir}} (T,L)\simeq L^{-d}X(\kappa tL^{1/\nu}),
\end{equation}
where $X(x)$ is an universal function and $\kappa$ is a
nonuniversal metric factor. In the case of SR interaction the
universal scaling structure of Eq.(\ref{CFH}) has been confirmed
by RG and exact calculations \cite{brankov00,krech94}.

The investigation of systems with LR interaction possesses some
peculiarities: due to the
 character of the interaction there exists a natural attraction between the surfaces bounding
 the system. In this case only results  within the spherical
 model with p.b.c. are obtained \cite{CD04}.
It is
 possible to formulate  some general statements:
the Casimir force is always negative, (i.e. it is a force of
attraction
 between the surfaces bounding the system) for any T and $\sigma\geq1$,
as well as for any  $T\geq T_{c}$ if $\sigma<1$; the behavior of
the Casimir force depends strongly on the range of interaction
$\sigma$, being (at $T=T_{c}$) a monotonically increasing function
of $\sigma$; in the neighborhood of  $T=T_{c}$, the Casimir force
is an increasing function of T if $\sigma>1$, and possesses a
complex behavior if $\sigma\leq1$.

For the Casimir force
 there is an exact expression obtained in the
framework of the quantum spherical model (see \cite{brankov00})
\begin{equation}
F_{{\rm Casimir}}(T,\lambda;L)\simeq L^{-(d+z)}X_{{\rm
Casimir}}(x,\rho ), \label{eq12}
\end{equation}
with scaling variable $x=L^{1/\nu }\left(\lambda^{-1} -\lambda
_c^{-1}\right), \ \mbox{and} \  \rho =L^z/L_\tau,  L_\tau =\lambda
/T.$ Here
  $\lambda $ is the quantum parameter that governs the
  transition near the quantum critical point $\lambda _c$.

The obtained exact expression for the {\it universal} scaling
function $X_{{\rm Casimir}} (x_,\rho )$ gives the possibility of
analysis including issues as the sign of the Casimir force and its
monotonicity  as a function of the temperature and parameter
 $\sigma$ \cite{CDT00}.

The author thanks J. Brankov, H. Chamati and D. Danchev for
valuable discussions. Support by Bulgarian NSF, Grant F-1402, is
acknowledged.

\bigskip
\end{document}